\documentclass[lettersize,journal]{IEEEtran}
\IEEEoverridecommandlockouts

\usepackage{cite}
\usepackage{amsmath,amssymb,amsfonts,amsthm}
\usepackage{mathtools}
\usepackage{algorithm}
\usepackage{algorithmic}
\usepackage{graphicx}
\usepackage{textcomp}
\usepackage{xcolor}
\usepackage{gensymb}
\usepackage{verbatim}
\usepackage{float}             
\usepackage{subfig}               
\usepackage{overpic} 
\def\BibTeX{{\rm B\kern-.05em{\sc i\kern-.025em b}\kern-.08em T\kern-.1667em\lower.7ex\hbox{E}\kern-.125emX}}

\renewcommand{\IEEEQED}{\IEEEQEDopen}

\begin{document}
\bstctlcite{IEEEexample:BSTcontrol}

\title{Pseudo MIMO (pMIMO): An Energy and Spectral Efficient MIMO-OFDM System}
%
\author{
        Sen Wang,
        Tianxiong Wang,
        Shulun Zhao,
        Zhen Feng,
        Guangyi Liu,\\
        Chunfeng Cui,
        Chih-Lin I,~\IEEEmembership{Life Fellow, IEEE,}
        and Jiangzhou Wang,~\IEEEmembership{Fellow, IEEE}
\thanks{S. Wang, T. Wang, S. Zhao, Z. Feng, G. Liu, C. Cui, C. L. I are with the China Mobile Research Institute, Beijing, China. (e-mail: \mbox{wangtianxiong@139.com}).}
\thanks{J. Wang is with the School of Engineering, University of Kent, Canterbury CT2 7NZ, U.K. He is also with the China Mobile Research Institute, Beijing, China. (e-mail: j.z.wang@kent.ac.uk).}
\vspace{-2em}}

\maketitle

\begin{abstract}
This article introduces an energy and spectral efficient multiple-input multiple-output orthogonal frequency division multiplexing (MIMO-OFDM) transmission scheme designed for the future sixth generation (6G) wireless communication networks. The approach involves connecting each receiving radio frequency (RF) chain with multiple antenna elements and conducting sample-level adjustments for receiving beamforming patterns. The proposed system architecture and the dedicated signal processing methods enable the scheme to transmit a bigger number of parallel data streams than the number of receiving RF chains, achieving a spectral efficiency performance close to that of a fully digital (FD) MIMO system with the same number of antenna elements, each equipped with an RF chain. We refer to this system as a ``pseudo MIMO'' system due to its ability to mimic the functionality of additional invisible RF chains. The article begins with introducing the underlying principles of pseudo MIMO and discussing potential hardware architectures for its implementation. We then highlight several advantages of integrating pseudo MIMO into next-generation wireless networks. To demonstrate the superiority of our proposed pseudo MIMO transmission scheme to conventional MIMO systems, simulation results are presented. Additionally, we validate the feasibility of this new scheme by building the first pseudo MIMO prototype. Furthermore, we present some key challenges and outline potential directions for future research.

\end{abstract}



\section{Introduction}
Over the last three decades, significant progress has been made in the advancement of multiple-input multiple-output (MIMO) technology. This innovative approach
has revolutionized the development of wireless communication technologies \cite{foschini1998limits, viswanathan2014past, larsson2014massive}. MIMO technology has played an increasingly important role in fourth generation (4G) and fifth generation (5G) communication systems, with the number of radio frequency (RF) chains 
at the base station (BS) increasing from 4 chains in 4G to 64 chains in 5G, and the massive MIMO technology has been successfully implemented in 5G \cite{chataut2020massive}. Although the 5G communication systems have effectively achieved the targets of a 3-fold increase in spectral efficiency and a 10-fold enhancement in energy efficiency, the focus of both academia and industry is on shifting to the evolution of MIMO for the forthcoming sixth generation (6G) wireless communications era \cite{dang2020should}. With the increasing frequency bands for future wireless communications, more antenna elements can fit in the same antenna aperture for both BSs and user terminals in 6G. While MIMO continues being one of the key enabling technologies in 6G wireless communications, and numerous research efforts have been devoted to design more sophisticated transceivers, deploying such systems in practice is still a formidable challenge.

\begin{figure*}[t]
	\centering
	\subfloat[]{\includegraphics[width=0.9\columnwidth]{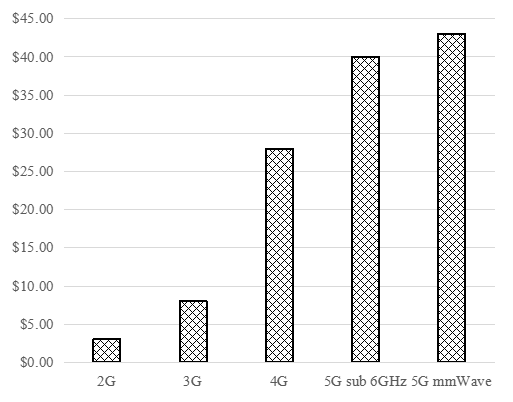}}\hspace{4pt}
	\subfloat[]{\includegraphics[width=0.9\columnwidth]{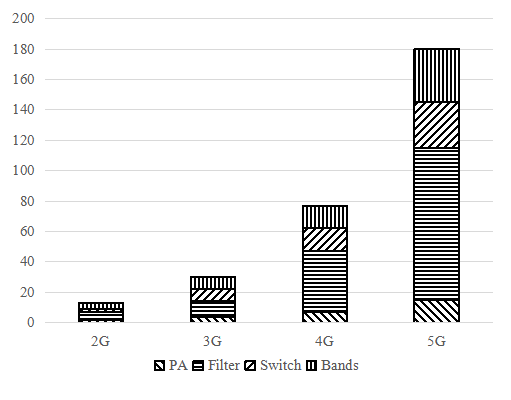}}\\
	\caption{Average RFFE cost and component quantity in mobile devices per generation: (a) average RFFE cost, and (b) average RFFE component quantity.}
\label{fig:RFFE evolution}
\end{figure*}

A notable concern associated with the MIMO technology is the substantial rise in hardware costs. The integration of additional RF chains into future MIMO systems inevitably leads to an increase in manufacturing expenses. The increase in hardware costs poses a significant challenge to the economic viability of deploying advanced MIMO technologies on a global scale \cite{andrews2014will}. In particular, the integration of more RF chains into mobile devices proves to be a complicated task in terms of chip manufacturing. The growing demand for sophisticated RF chips necessitates more advanced manufacturing processes, consequently driving up hardware costs.
This escalation in costs, in turn, directly impacts on the pricing of devices. As shown in Fig.~\ref{fig:RFFE evolution}a, the average cost of radio frequency front-end (RFFE) in mobile devices increases with the development of each generation of wireless communications. Therefore, the evolution of MIMO technology has to carefully consider the balance between performance enhancements and economic feasibility.

Another key obstacle impeding the development of MIMO technology is the increasingly pronounced power consumption dilemma. The RF chain consists of a series of active components that collectively enable the transmission and reception of radio signals. Key components of an RF chain include power amplifiers (PA), mixers, filters, etc. As shown in Fig.~\ref{fig:RFFE evolution}b, the quantity of these components is increasing with the development of wireless communications.
Power consumption concern of MIMO technology is due to the active nature of these components. Incorporating a greater number of RF chains in future BSs and user terminals will result in a substantial surge in power consumption \cite{prasad2017energy}. This heightened energy demand not only poses environmental sustainability challenges but also raises operational costs for the deployment and maintenance of communication networks. 


Furthermore, existing MIMO transmission schemes, which take into account the balance among hardware costs, power consumption, and communication performance, still have untapped potential for further enhancement. Specifically, the hybrid digital and analog MIMO architecture, where each RF chain is connected to multiple antenna elements, has attracted extensive research interests \cite{gao2016energy}. This MIMO configuration significantly reduces hardware complexity and power consumption at the cost of certain capacity loss. 
Nevertheless, it has to acknowledge that the conventional hybrid MIMO architecture is a trade-off plan that does not fully exploit spatial resources. The maximum number of data streams that can be successfully transmitted in a hybrid MIMO system is determined by the lesser of the two: the number of transmitting RF chains and the number of receiving RF chains. Consequently, the spectral efficiency of a hybrid MIMO system falls below that of a fully digital (FD) MIMO system with the same number of antenna elements, each equipped with an RF chain. The question of whether the conventional hybrid MIMO system can be transformed in such a way that its spectral efficiency reaches that of the FD MIMO counterpart without consuming additional power remains open and unresolved.


Motivated by the above challenges impeding the advancement of MIMO technology, this article proposes a novel transceiver architecture and a corresponding transmission scheme for MIMO-OFDM systems. The innovation involves a high-speed reconstructing antenna system on the receiver side. By rapidly reconstructing the beamforming patterns of the antennas and upsampling the OFDM symbols on the receiver,
the technology enables the transmission of a greater number of parallel data streams than the available RF chains. The system, termed ``pseudo MIMO'', operates as if it has additional ``virtual'' RF chains. 
In contrast to the conventional hybrid MIMO system, the proposed pseudo MIMO system attains higher spectral efficiency while maintaining similar hardware complexity and power consumption. On the other hand, when compared with the FD MIMO system, the pseudo MIMO system achieves a comparable spectral efficiency with significantly reduced hardware costs and power consumption.
In the following subsections, we first introduce the principles and the hardware architecture of the proposed pseudo MIMO system. Then, the advantages and potential application scenarios of pseudo MIMO technology are presented. Subsequently, the simulation results and the experimental findings from the prototype testing are presented. Finally, some future research directions and challenges for pseudo MIMO are discussed.

\section{Principles and Hardware Architecture}
In this section, we present the principles and hardware architecture of the pseudo MIMO communication systems. We begin with detailing the transmission scheme of a pseudo MIMO system with two transmitting RF chains and one receiving RF chain. Following that, we highlight that the proposed scheme can be generalized to systems with any number of RF chains. Lastly, we discuss the hardware implementation of pseudo MIMO in practice.

\subsection{Physical Principles of Pseudo MIMO}
\begin{figure*}[t]
	\centering
	\subfloat[]{\includegraphics[width=.97\columnwidth]{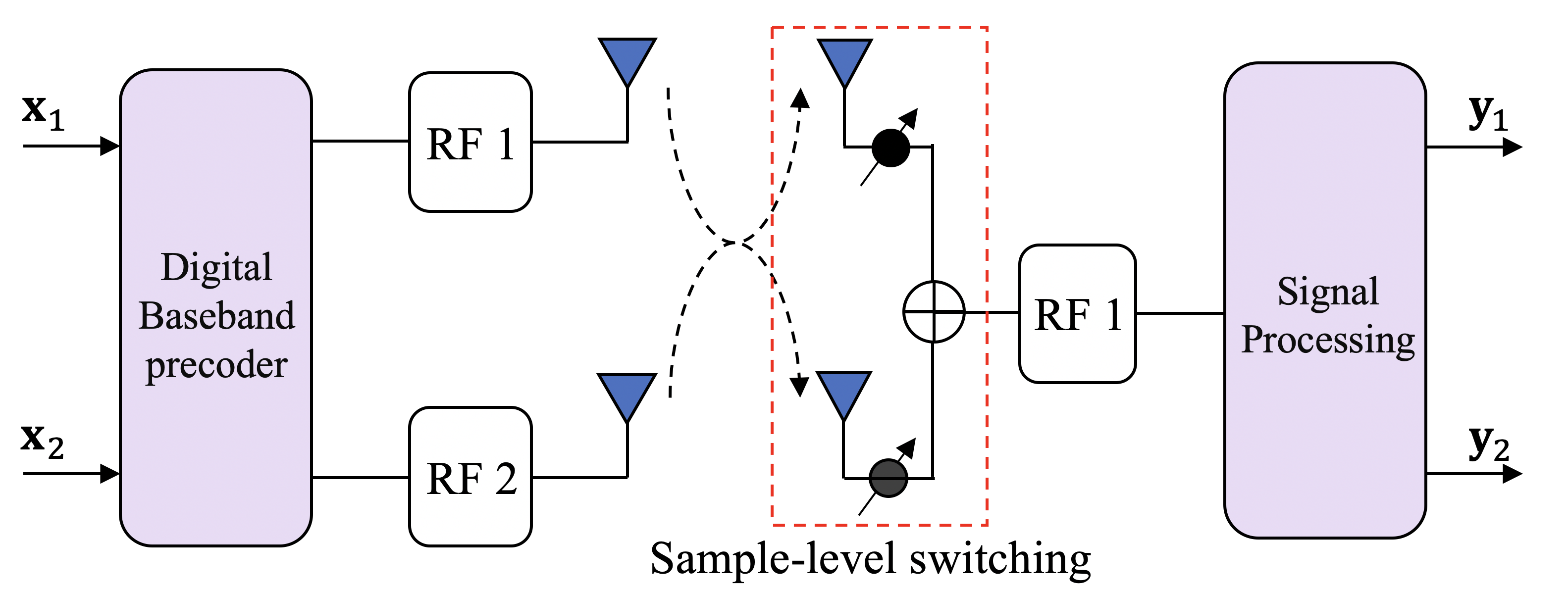}} \hspace{8pt}
	\subfloat[]{\includegraphics[width=.97\columnwidth]{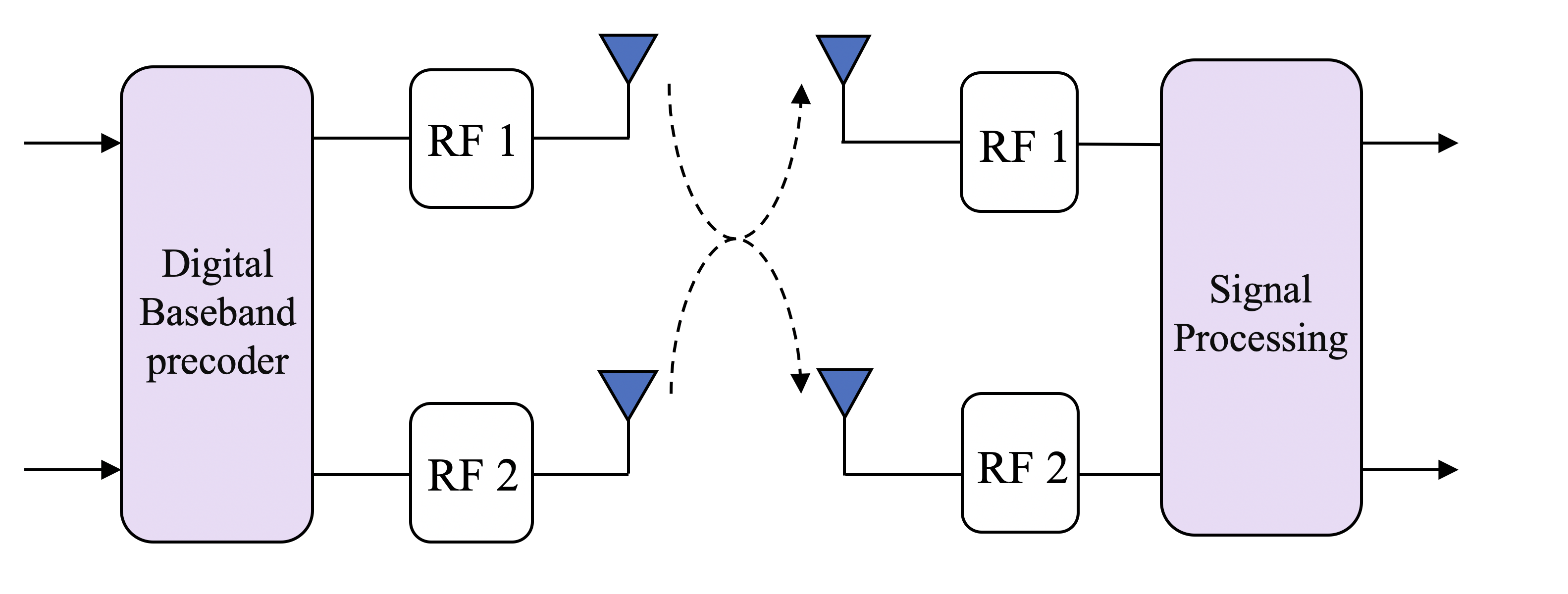}}\\
	\subfloat[]{\includegraphics[width=.97\columnwidth]{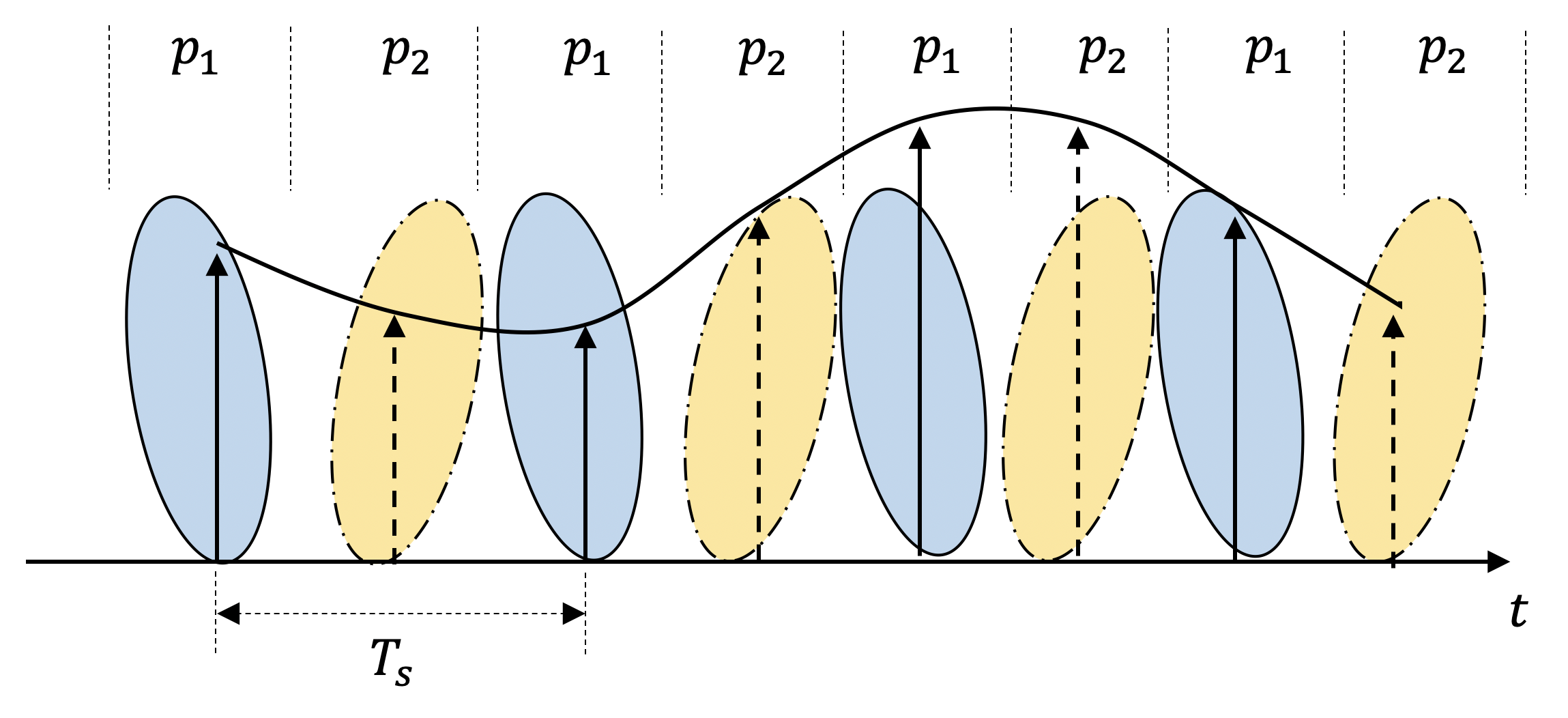}} \hspace{8pt}
	\subfloat[]{\includegraphics[width=.97\columnwidth]{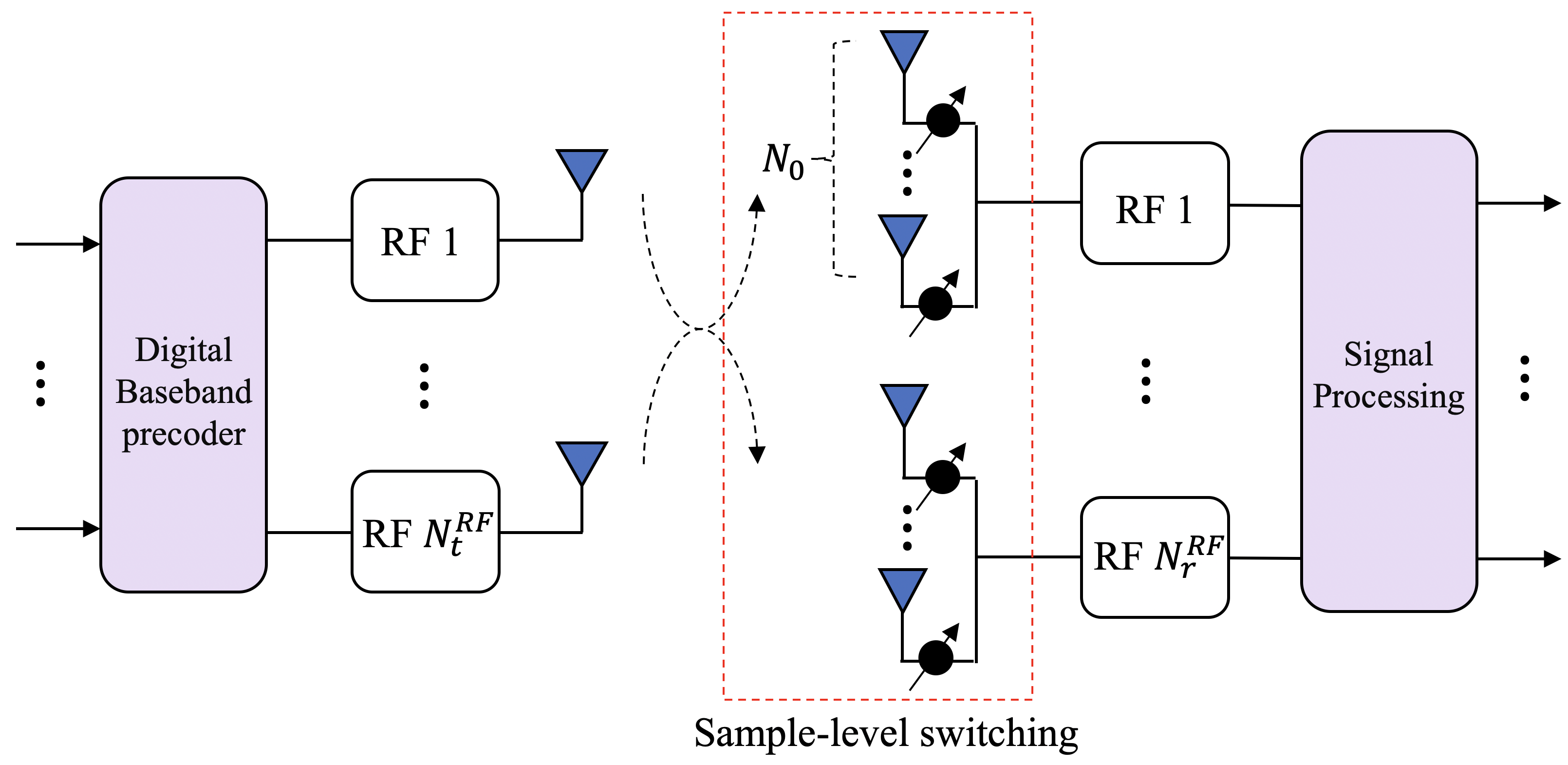}}
	\caption{Schematic diagram and physical principle of pseudo MIMO systems: (a) structure of a pseudo MIMO system with two transmitting RF chains and one receiving RF chain, (b) structure of a conventional FD MIMO system with two transmitting RF chains and two receiving RF chains, (c) illustration of sample-level receiving beamforming patterns reconstructing and upsampling, and (d) structure of a general pseudo MIMO system.}
\label{principles11}
\end{figure*}

Fig.~\ref{principles11}a depicts the schematic diagram of a pseudo MIMO communication system. The transmitter is equipped with two RF chains, each incorporating one antenna element. The receiver is similar to a hybrid MIMO architecture. The receiver features a single RF chain but is equipped with two antenna elements. In the context of conventional MIMO framework, this system allows for the transmission of only one data stream, owing to the single RF chain at the receiver. Consequently, its communication performance falls below that of its counterpart with the same number of antenna elements but two RF chains, as shown in Fig.~\ref{principles11}b. However, with the proposed pseudo MIMO scheme, if the beamforming patterns of the receiving antenna elements can be varied rapidly per OFDM sample, the system presented in Fig.~\ref{principles11}a can support a maximum of two data streams transmissions simultaneously without the need for extra RF chains. 

The transmission scheme of a pseudo MIMO communication system at the transmitter side is similar to a conventional MIMO-OFDM system. As shown in Fig.~\ref{principles11}a, two distinct OFDM symbols, each of length $n$ and denoted as $\mathbf{x}_1$ and $\mathbf{x}_2$, are transmitted through the two RF chains of the transmitter. These two frequency domain OFDM symbols undergo an inverse fast Fourier transform (IFFT) operation, transforming them into the corresponding time domain digital signals $\hat{\mathbf{x}}_1$ and $\hat{\mathbf{x}}_2$. Both $\hat{\mathbf{x}}_1$ and $\hat{\mathbf{x}}_2$ consist of $n$ samples, and it is assumed that the time duration between two adjacent samples is $T_s$. Following the IFFT operation, the time-domain digital signals go through a digital-to-analog converter (DAC), transforming them into analog signals before transmission through the wireless channels via the transmitting antenna elements.

At the receiver side, the beamforming patterns of antenna elements dynamically reconstruct between two different patterns, denoted as $p_1$ and $p_2$, within each OFDM sampling period, effectively capturing the same OFDM symbols with different beamforming configurations. As illustrated in Fig. \ref{principles11}c, a pair of beamforming patterns is applied within the time duration of one original sampling period, $T_s$. The OFDM symbols received by two antenna elements are combined into a single chain before a crucial step—upsampling by a ratio of 2 through an analog-to-digital converter (ADC). The upsampling process is a key step in pseudo MIMO transmission scheme. In the time domain, upsampling serves as a form of interpolation, inserting extra samples based on existing ones. Essentially, upsampling strategically introduces redundant information into the original OFDM symbols without consuming additional time and frequency resources. The receiver utilizes this redundancy to recover the two different symbols. Through the process of upsampling, the received signals corresponding to the beamforming pattern $p_1$ can be distinguished from the signals received by the beamforming pattern $p_2$. For example, in the particular case shown in Fig. \ref{principles11}c, samples with odd indices, represented by solid lines, are captured by $p_1$, while those with even indices, obtained through upsampling and represented by dashed lines, are captured by $p_2$. Following the upsampling operation, the combined digital symbol is split into two symbols: one comprising only the odd-index elements, and the other containing only the even-index elements. Independent fast Fourier transform (FFT) operations are applied to the two separated symbols, resulting in two different frequency domain OFDM symbols, i.e., $\mathbf{y}_1$ and $\mathbf{y}_2$. Representing the entire signal processing and wireless transmission procedures by a matrix $\mathbf{H}$, it can be proved that $\mathbf{H}$ is full-rank as long as two different receiving beamforming patterns are employed. Decoding of $\mathbf{x}_1$ and $\mathbf{x}_2$ can be accomplished using conventional methods such as zero forcing (ZF), minimum mean-square error (MMSE), etc. Hence, the system has the capability to facilitate the transmission of two parallel data streams, even when the receiver has only one RF chain.

It is worth noting that the pseudo MIMO transmission scheme is applicable to systems with more antennas. A general pseudo MIMO communication system is illustrated in Fig.~\ref{principles11}d. The transmitter comprises $N_t^{RF}$ RF chains, each with one antenna element. The receiver has $N_r^{RF}$ RF chains, each with $N_0$ antenna elements, resulting in a total of $N_r^{A} = N_r^{RF} N_0$ receiving antenna elements. In contrast to conventional MIMO systems limited to $\min\{N_t^{RF}, N_r^{RF}\}$ data streams transmissions, the pseudo MIMO setup can support up to $\min\{N_t^{RF}, N_r^{A}\}$ parallel data streams transmissions. Assuming that $N_r^A$ is less than $N_t^{RF}$, pseudo MIMO is capable of increasing the number of parallel data streams by $N_0$ times. To accomplish this, the beamforming patterns of the antenna elements on the receiver side need to dynamically switch among $N_0$ different patterns within one original sampling period, $T_s$. Additionally, the received analog signal must be upsampled by a ratio of $N_0$. Thus a full-rank equivalent channel can be established. Remarkably, the $N_r^{A}$ distinct OFDM symbols can be successfully decoded with just $N_r^{RF}$ RF chains. With the pseudo MIMO transmission scheme, the maximum number of data streams that can be transmitted is determined by the number of receiving antenna elements, rather than the receiving RF chains. Therefore, the proposed pseudo MIMO system remarkably improves spectral efficiency in comparison to the hybrid MIMO. Moreover, through the optimal design of receiving beamforming patterns, the achievable performance of the pseudo MIMO system can closely approach that of the conventional FD MIMO, representing the theoretical performance upper bound.


\subsection{Potential Hardware Designs of Pseudo MIMO}
Implementing the pseudo MIMO transmission scheme in future wireless communication networks requires further investigations, specifically in achieving rapid reconstructing of beamforming patterns at the receiver. A promising solution is to integrate an analog phase shifter into each antenna element. These analog phase shifters enable the alteration of phase shifts, realizing different beamforming patterns. It should be noted that the analog phase shifters should be able to rapidly switch among different values within the duration of an OFDM sample. In terms of power consumption, the application of analog phase shifters in both pseudo MIMO and hybrid MIMO results in comparable power consumption.

\begin{figure*}[t]
    \centerline{\includegraphics[width=1.5\columnwidth]{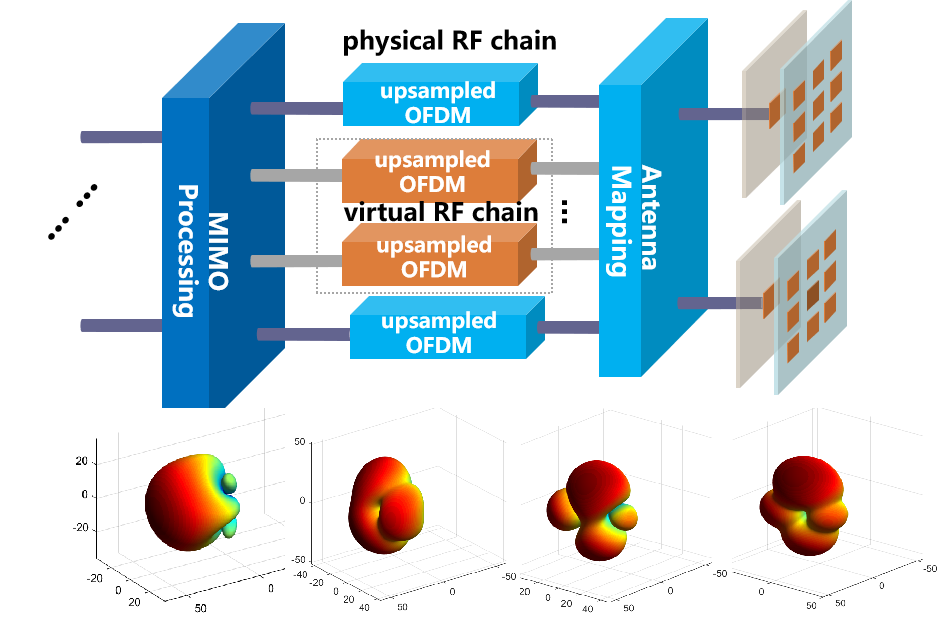}}   
    \caption{Schematic Diagram of a pseudo MIMO receiver equipped with reconfigurable antenna elements.}
    \label{recon_sys}
\end{figure*}

Another promising solution is to apply the ``reconfigurable antenna'' technology in pseudo MIMO. The fundamental idea of a reconfigurable antenna is to alter the electromagnetic (EM) properties of an antenna artificially \cite{ying2023reconfigurable}. A reconfigurable antenna can be implemented using various technologies, such as semiconductors, liquid metal, graphene, and liquid crystal. Fig.~\ref{recon_sys} illustrates the implementation of a pseudo MIMO receiver using reconfigurable antenna technology. In this setup, each physical RF chain corresponds to multiple reconfigurable antenna elements, allowing the creation of ``virtual RF chains''. Furthermore, integrating reconfigurable antennas into a pseudo MIMO system will not lead to a significant increase in power consumption. For instance, a PIN diode based reconfigurable antenna demonstrates low power consumption, with the power requirement of a PIN diode being less than 1\,mW \cite{tang2020mimo}, significantly lower than that of an analog phase shifter.

\section{Advantages and Applications of Pseudo MIMO in Wireless Networks}
Having presented the underlying physical principles and hardware designs of pseudo MIMO, this section delves into the prospective applications and potential advantages of applying pseudo MIMO in future wireless communications.

\subsection{Reduced Energy Consumption and Hardware Cost}


In contrast to the conventional FD MIMO systems, the pseudo MIMO technology holds a great promise in reducing the number of RF chains in mobile devices. RF chains constitute a significant source of both energy consumption and hardware costs for mobile devices. The continuous expansion of frequency bands, carrier aggregation, and the implementation of high-order modulation technologies have resulted in a notable surge in the number and complexity of RF chains. The escalating cost of RF chains leads to an increase in device prices. Moreover, the number of RF chains directly influences the power consumption of the device. The proposed pseudo MIMO system effectively mitigates both the cost and energy consumption of devices, since it can achieve spectral efficiency comparable to that of a conventional FD MIMO system by using a reduced number of RF chains.
\subsection{Enhanced Spectral Efficiency}


Compared with the conventional hybrid MIMO systems, the key advantage of integrating pseudo MIMO technology into future wireless networks is its ability to enhance spectral efficiency. The pseudo MIMO systems fully exploit spatial resources, resulting in a significant boost to spectral efficiency. This improvement stems from the dynamic reconfiguration of equivalent channels on a sample-by-sample basis. Through the sequential switching of receiving beamforming patterns at each sample, different groups of these OFDM time domain samples can experience different equivalent channels. The notable enhancement in spectral efficiency also yields additional benefits, outlined as follows.

\subsubsection{Higher Sensing Resolution}


The pseudo MIMO technology is expected to enhance the sensing performance of the integrated sensing and communications (ISAC) application in future 6G wireless networks. ISAC is envisioned to play a vital role in future wireless communications, with sensing accuracy and capture range serving as two key performance metrics \cite{liu2022integrated}. Theoretically, sensing resolution is directly proportional to the spectral efficiency, while the capture range is inversely proportional to the minimum time unit. Pseudo MIMO technology offers a dual advantage in enhancing sensing performance. Compared with the hybrid MIMO architecture, pseudo MIMO enhances spectral efficiency, thereby elevating sensing resolution. Simultaneously, the inherent upsampling process is beneficial for expanding the capture range.

\subsubsection{Expanded Wireless Coverage}

The pseudo MIMO technology is also capable of enhancing the coverage area of future wireless networks. As the deployment frequency bands continue to increase, cell coverage performance tends to degrade \cite{alsaedi2023spectrum}. Taking China Mobile as an example, the current main commercial frequency band for the 5G network is around 2.6\,GHz. According to the World Radiocommunication Conference 2023 (WRC-23) regulations, the main commercial frequency band for the future 6G network will be around 7\,GHz. However, the increase in frequency bands also introduces opportunities for incorporating a greater number of antenna elements within the same antenna aperture. To tackle the issue of inadequate wireless coverage in higher frequency bands, applying the pseudo MIMO technology proves beneficial due to its capability for increasing spectral efficiency.


\subsection{Convenient Deployment and Strong Compatibility}

The pseudo MIMO scheme exhibits strong backward and forward compatibility, and can be integrated into future wireless networks seamlessly. Legacy devices can still work using the conventional MIMO hardware architecture. Simultaneously, the pseudo MIMO scheme can be flexibly applied on new devices, as long as the device capabilities and channel state information (CSI) are reported to the BS.
This adaptability ensures that existing devices continue to operate effectively, while new devices can benefit from the advantages of the pseudo MIMO scheme based on their unique capabilities. Notably, the proposed scheme facilitates the increase of single-user MIMO transmission layers, thereby mitigating the demand for complex multi-user MIMO pairing.

\section{Numerical and Experimental Results}

In this section, numerical results obtained through Monte Carlo simulations are presented. We compare the performance of the proposed pseudo MIMO system with that of the conventional MIMO systems. Additionally, the first pseudo MIMO prototype is built, validating the feasibility of the pseudo MIMO transmission scheme through prototype testing.

\subsection{Performance Comparison with Conventional MIMO}
The simulations utilize a multipath channel model with 15 paths, where the fading of each path follows the Rayleigh distribution. We denote a pseudo MIMO and a hybrid MIMO system with $N_t^{RF}$ transmitting RF chains and $N_r^{RF}$ receiving RF chains, with each receiving RF chain equipped with $N_0$ antenna elements as ``pMIMO $N_t^{RF} \times N_r^{RF} \times N_0$'' and ``hybrid MIMO $N_t^{RF} \times N_r^{RF} \times N_0$'', respectively. On the other hand, a conventional FD MIMO system with $N_t^{RF}$ transmitting RF chains and $N_r^{RF}$ receiving RF chains is labeled as ``FD MIMO $N_t^{RF} \times N_r^{RF}$''. 

\begin{figure*}[t]
	\centering
	\subfloat[]{\includegraphics[width=.65\columnwidth]{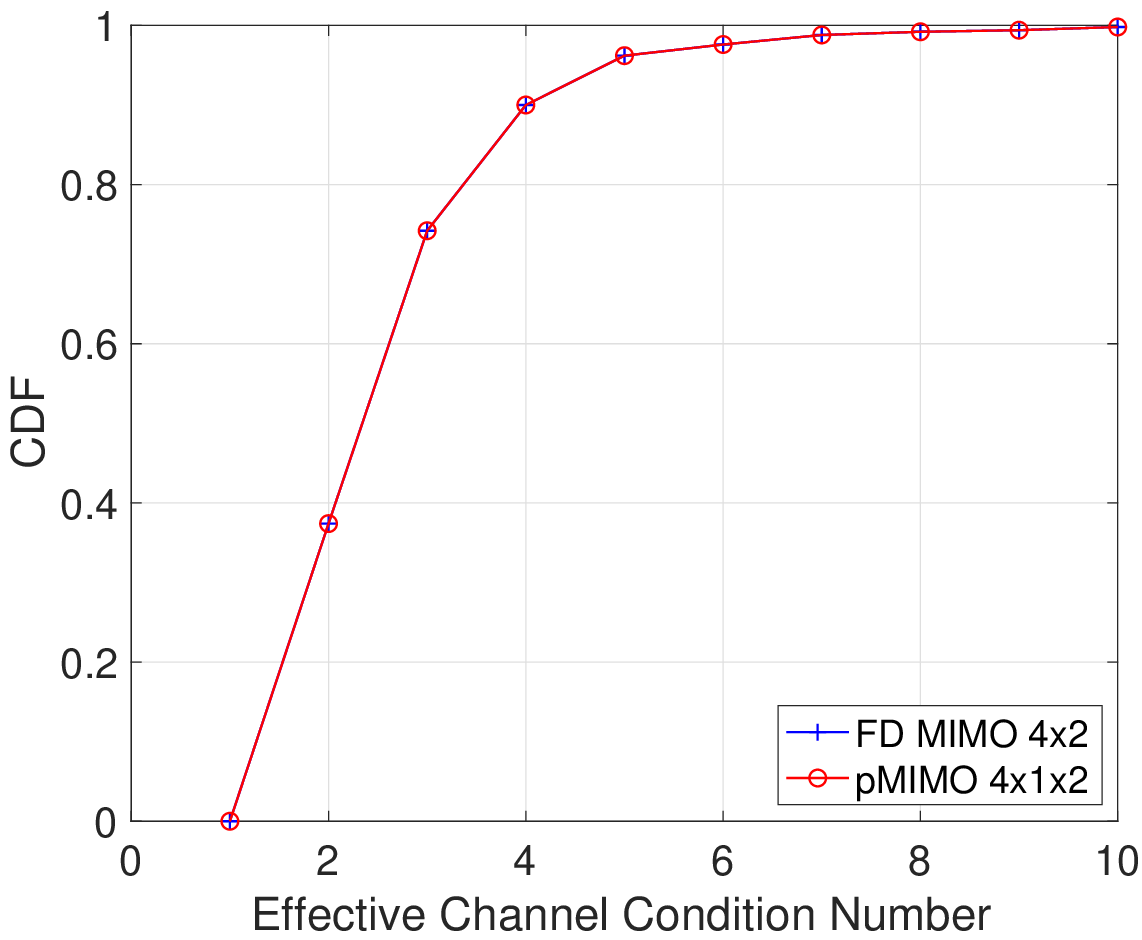}}\hspace{8pt}
	\subfloat[]{\includegraphics[width=.65\columnwidth]{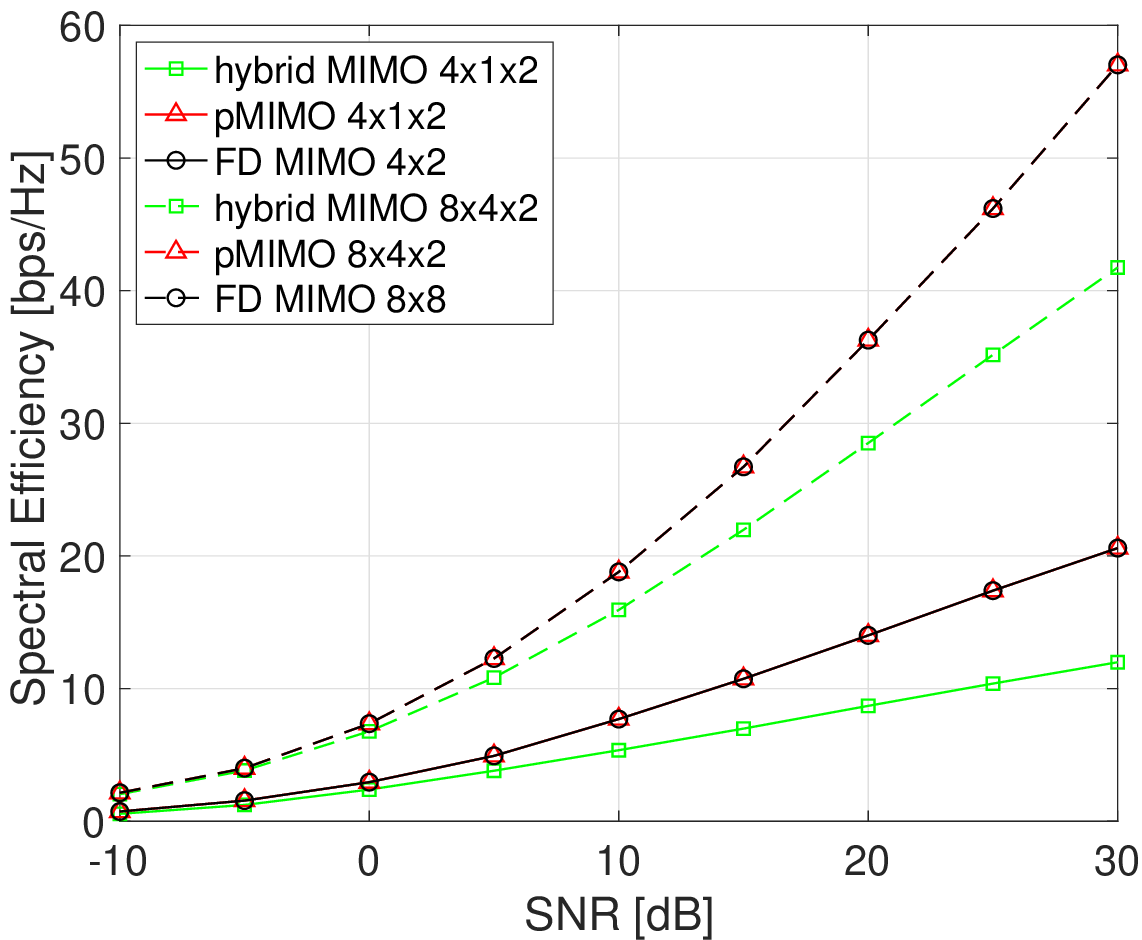}}\hspace{8pt}
    \subfloat[]{\includegraphics[width=.65\columnwidth]{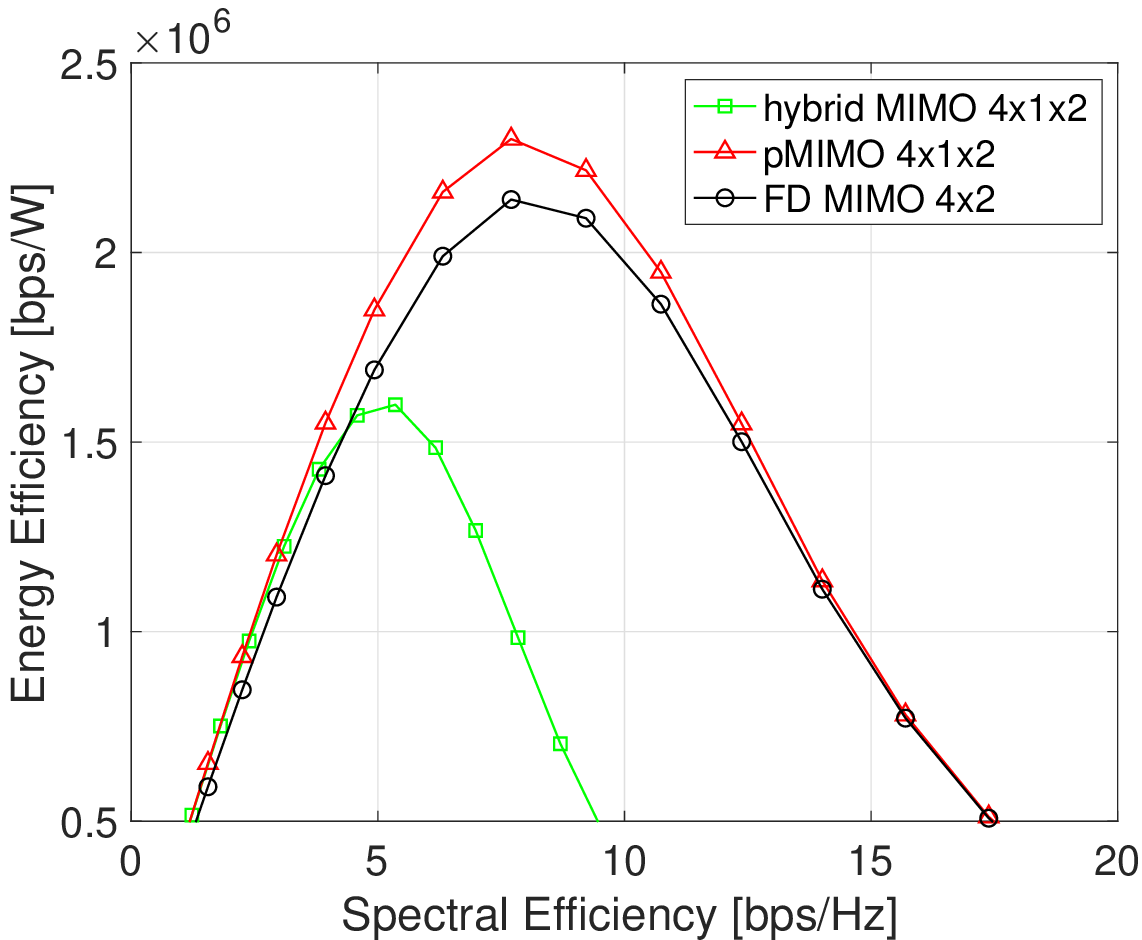}}
	\caption{Performance evaluation of pseudo MIMO: (a) CDF of effective channel condition numbers, (b) spectral efficiency comparison among pseudo MIMO with optimal receiving beamforming, conventional FD MIMO and hybrid MIMO, and (c) energy efficiency-spectral efficiency relationship.}
\label{fig:pMIMO perf eval}
\end{figure*}

Fig.~\ref{fig:pMIMO perf eval}a illustrates the cumulative distribution function (CDF) of the effective channel condition numbers for both pseudo MIMO and FD MIMO systems. The effective channel condition number is defined as the ratio of the largest eigenvalue to the smallest eigenvalue of the channel matrix in a MIMO system. A lower effective channel condition number indicates a more favorable channel condition for transmitting multiple parallel data streams.
Both pseudo MIMO and FD MIMO systems are equipped with two receiving antenna elements. Notably, the pseudo MIMO system has only one receiving RF chain, while the FD MIMO system utilizes two receiving RF chains. Despite this difference, the results show that the effective channel condition number of the pseudo MIMO system closely aligns with that of the FD MIMO system. This phenomenon suggests that, due to the receiver design and signal processing methods, the capability of the pseudo MIMO system to support parallel data streams transmissions is very similar to that of the FD MIMO system, even with fewer RF chains.

Fig.~\ref{fig:pMIMO perf eval}b compares the spectral efficiency among pseudo MIMO, conventional FD MIMO and hybrid MIMO with varying antenna configurations. Pseudo MIMO configurations vary in $N_t^{RF} \times N_r^{RF} \times N_0 \in \{4 \times 1 \times 2,  8 \times 4 \times 2\}$, while FD MIMO configurations vary in $N_t^{RF} \times N_r^{RF} \in \{4 \times 2, 8 \times 8\}$. Hybrid MIMO configurations vary in $N_t^{RF} \times N_r^{RF} \times N_0 \in \{4 \times 1 \times 2,  8 \times 4 \times 2\}$. With optimal receiving beamforming patterns, pseudo MIMO significantly enhances spectral efficiency compared to hybrid MIMO with the same number of receiving RF chains. For instance, a $4 \times 1 \times 2$ pseudo MIMO achieves $13 {\rm\,bps/Hz}$ compared to $9 {\rm\,bps/Hz}$ for a $4 \times 1 \times 2$ hybrid MIMO when the transmit SNR is 20\,dB, indicating almost a $50\,\%$ performance gain. Additionally, the spectral efficiency of the pseudo MIMO closely aligns with that of the FD MIMO, even when the FD MIMO has more RF chains, namely, the $4 \times 1 \times 2$ pseudo MIMO compared to the $4 \times 2$ FD MIMO, and the $8 \times 4 \times 2$ pseudo MIMO compared to the $8 \times 8$ FD MIMO. This highlights the potential of the pseudo MIMO in reducing the required RF chains.

Fig.~\ref{fig:pMIMO perf eval}c depicts the relationship between energy efficiency and spectral efficiency, considering both transmit power and static power from hardware components. The curve is non-monotonic due to static power consumption, modeled based on \cite{han2015large}. Assuming negligible power difference between an analog phase shifter and a reconfigurable antenna element, the pseudo MIMO system exhibits the highest energy efficiency. Fig.~\ref{fig:pMIMO perf eval}c reveals that, in contrast to both hybrid MIMO and FD MIMO, the pseudo MIMO system has an advantage in reducing power consumption while achieving the same spectral efficiency.


    

%
\subsection{Pseudo MIMO Prototype Study}
\begin{figure*}[t]
	\centering
	\subfloat[]{\includegraphics[width=1.2\columnwidth]{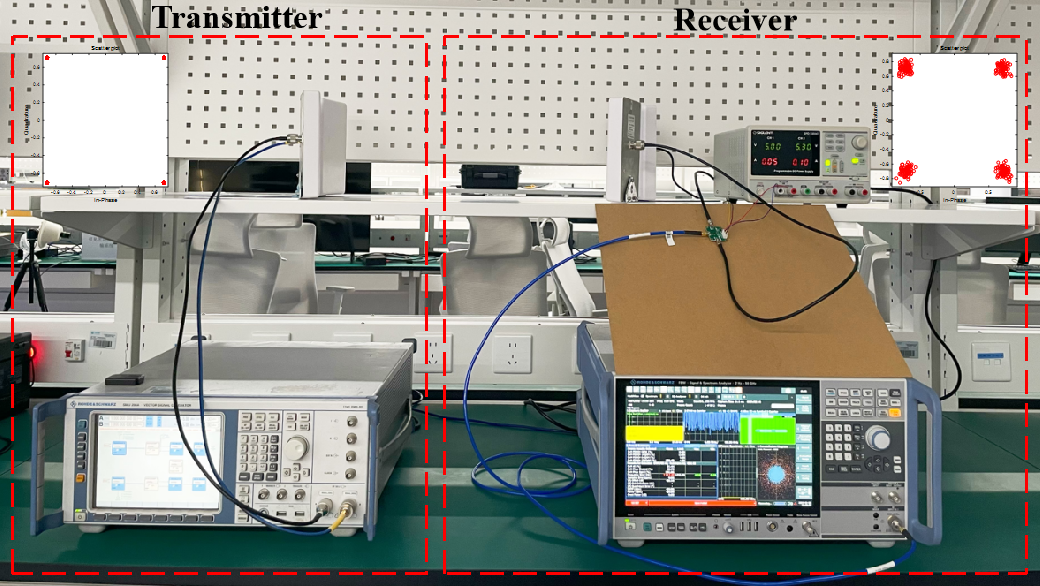}} \hspace{8pt}
	\subfloat[]{\includegraphics[width=0.7\columnwidth]{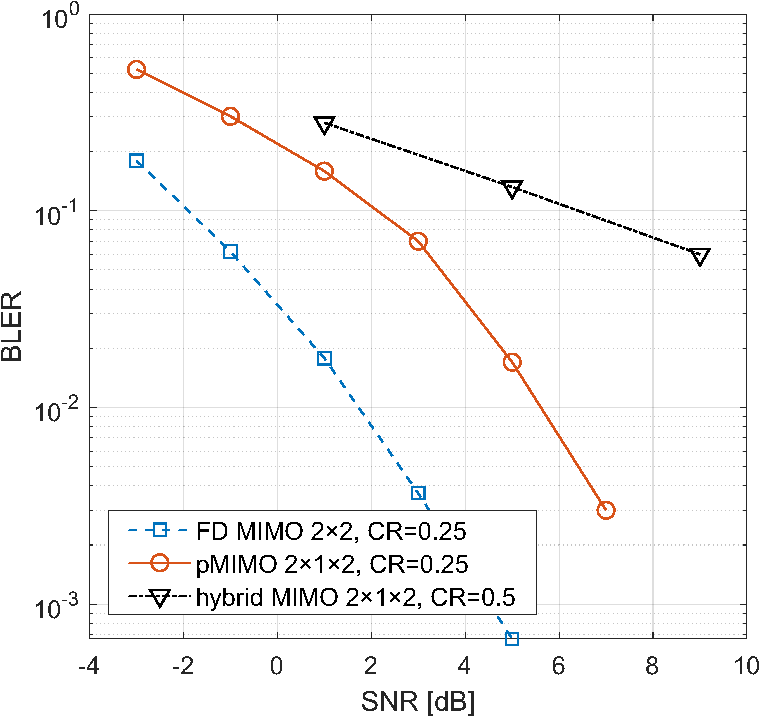}}
	\caption{Pseudo MIMO prototype: (a) structure of the pseudo MIMO prototype and the transmitted and received constellation symbols, and (b) BLER performance of the pseudo MIMO system.}
\label{prototype}
\end{figure*}

The pseudo MIMO prototype is illustrated in Fig.~\ref{prototype}a. On the left side, the transmitter of the pseudo MIMO system is equipped with two antenna elements. Both of these antenna elements are connected to a Rhode\&Schwarz (R\&S) signal generator SMW200A, which is equipped with two RF chains.
The receiver of the pseudo MIMO system, positioned on the right, incorporates two antenna elements. Each antenna element is connected to the respective input port of an RF switching device capable of altering the beamforming patterns of the receiving antenna elements. The single output port of the device connects to the single-port signal and spectrum analyzer R\&S FSW.

Two waveform files, each containing independent quadrature phase shift keying (QPSK) modulated symbol streams, are precoded before being imported into the R\&S signal generator. The signal generator upconverts the waveform files to the 2.4\,GHz frequency band, transmitting them from the two antenna elements. The beamforming patterns of the receiving antenna elements rapidly switch per OFDM sample under the control of the high-speed RF device. The combined single data stream output is processed by the R\&S signal and spectrum analyzer for upsampling and data acquisition.
Through the rapid switching of beamforming patterns and the implemented signal processing procedures, the pseudo MIMO hardware prototype successfully decodes two independent data streams, achieving the transmission of two parallel data streams with only one receiving RF chain in the signal and spectrum analyzer.

Fig.~\ref{prototype}b presents the block error rate (BLER) of the pseudo MIMO prototype, in comparison with both FD MIMO and hybrid MIMO configurations: $2 \times 2$ FD MIMO transmitting two parallel data streams, and $2 \times 1 \times 2$ hybrid MIMO transmitting one data stream. In the legend, ``CR'' denotes the channel coding rate. For fairness, the CR for the hybrid MIMO with one receiving RF chain is doubled to maintain equal overall code rates.
As expected, the BLER of the pseudo MIMO system is lower than that of the hybrid MIMO with only one receiving RF chain. However, since the beamforming patterns of the pseudo MIMO receiver are not optimized, and only two fixed patterns are employed and switched sequentially, its BLER is higher than the FD MIMO with two receiving RF chains.

\section{Challenges and Research Directions}
The pseudo MIMO scheme has the potential to become an important physical layer technique in future 6G wireless networks. Nevertheless, certain challenges need to be addressed before commercial application becomes feasible. Hereafter, we outline some key challenges and propose potential research directions to address them.

\subsection{Channel Estimation and CSI Feedback}

Accurate channel estimation and CSI feedback are critical for wireless communications \cite{6798744}. However, since the beamforming patterns rapidly switch within the duration of an OFDM symbol, channel estimation and CSI feedback become challenging for pseudo MIMO systems. The rapid switching of beamforming patterns may lead to inter-carrier interference (ICI), making accurate channel estimation unfeasible. Moreover, channel estimation of pseudo MIMO systems should span across all the different beamforming patterns, inevitably increasing training and feedback overhead. Therefore, it is important to design effective methods that mitigate the impact of ICI and make channel estimation and CSI feedback efficient.



Novel channel estimation approaches should be employed. The design of reference signal for channel estimation requires more sophisticated methods, such as compressed sensing and deep learning \cite{8353153}, which can estimate the channel parameters by exploiting the correlations between the beamforming patterns. Moreover, since multiple beamforming patterns are adopted by the receiver, CSI reporting should be comprehensive which can inform the channel quality under different conditions of beamforming patterns.
In addition, advanced receivers may be required to mitigate the ICI resulting from the switching of beamforming patterns in pseudo MIMO systems.
Techniques, such as interference rejection combining (IRC) equalizers, can be enhanced to eliminate ICI between individual subcarriers and suppress interference between layers. Furthermore, pseudo MIMO transmitters can estimate and preprocess ICI through appropriate digital precoding. This is particularly effective in downlink transmissions, where suppressing ICI at the transmitter can significantly reduce the computational load and power consumption of user terminals.

\subsection{Synchronization}

The synchronization process is important in wireless communication systems. In current systems, the requirement of the synchronization typically operates at the OFDM symbol level, locating boundaries of OFDM symbols. In pseudo MIMO systems, rapid reconstructing of receiving beamforming patterns demands synchronization precision at the OFDM sample level, making the time resolution of existing systems insufficient. One potential solution to enhance precision is expanding the bandwidth of the synchronization signals. Furthermore, since the correlation detection is done in the time domain, the synchronization capture range can be enhanced by placing the synchronization signals densely in the frequency domain. However, these approaches will lead to increased synchronization overhead, affecting the overall performance. Therefore, it is necessary to design the synchronization signals carefully, achieving an optimal trade-off between resource occupation and synchronization detection performance.

\subsection{High-Speed Reconfigurable Antenna Design}

In a pseudo MIMO system, the rapid switching of receiving beamforming patterns presents new challenges for the antenna elements design in mobile devices. As bandwidth increases, the sampling interval decreases, demanding faster variation of receiving beamforming patterns. For instance, achieving 100\,MHz working bandwidth requires the receiving beamforming patterns to switch every 4 nanoseconds. Unfortunately, existing analog phase shifters operate in the microsecond range, falling short of the requirements for pseudo MIMO systems. Consequently, one potential research direction involves optimizing the hardware design to construct a nanosecond-level high-speed analog phase shifter.
On the other hand, reconfigurable antennas can meet the demands of faster beamforming adjustments. A semiconductor based reconfigurable antenna, for instance, can achieve a minimum reconstructing interval as short as 3 nanoseconds, constrained by the limit of a PIN diode. However, a semiconductor based antenna can only achieve discrete receiving beamforming, and designing high precision beamforming makes the hardware design complicated. 
Conversely, opting for low precision beamforming design could restrict the potential performance gains of pseudo MIMO systems. Therefore, if a semiconductor based reconfigurable antenna is employed, careful consideration must be given to the trade-off between performance gains and hardware complexity.
Apart from semiconductor based reconfigurable antennas, graphene based reconfigurable antennas have the potential to support even picosecond-level beamforming adjustments. However, this direction remains less explored and deserves further research efforts.

\section{Conclusions}
An energy and spectral efficient MIMO-OFDM transmission scheme called pseudo MIMO was introduced in this article. We studied the physical principles and hardware architectures of this new technology, and explored its potential advantages. Notably, our simulation results and prototype testing validated the feasibility of pseudo MIMO, and demonstrated its capability to achieve the spectral efficiency close to that of a conventional FD MIMO system but with power consumption and hardware costs similar to a hybrid MIMO system. As the pseudo MIMO technology is still in its infancy, we discussed critical challenges associated with its implementation, and proposed research directions essential to making pseudo MIMO a key technology element for future wireless communications.

\bibliographystyle{IEEEtran}
\bibliography{irs}

\end{document}